\documentclass[conference]{IEEEtran}

\usepackage{cite}  

\usepackage[square, numbers, comma]{natbib} 
 \usepackage{hyperref}
\usepackage{amsmath}
\usepackage{amssymb, amsmath}
\usepackage{float}
\usepackage{graphicx}
\usepackage{flushend}
\usepackage{stmaryrd}
\usepackage{tabularx}
\usepackage[utf8]{inputenc}
\usepackage{ulem}
\usepackage{graphicx}
\usepackage{graphics}
\usepackage{comment}
\usepackage{multirow,rotating}
\usepackage{cancel} 
\usepackage{colortbl}
\usepackage{xcolor}
\usepackage{amsmath}
\usepackage{amsfonts} 
\usepackage{booktabs}
\usepackage{enumerate} 
\usepackage{array}
\usepackage[ruled]{algorithm2e}
\usepackage{mathtools}
\usepackage{pgfplots}
\usepackage{algorithmic}
\pgfplotsset{compat=1.14} 
\usepackage{color} 
\usepackage{tikz}    
\usepackage{subcaption}

\usepackage{amsmath}

\usepackage{enumerate}
\def\sqw{\hbox{\rlap{\leavevmode\raise.3ex\hbox{$\sqcap$}}$ \sqcup$}}
\def\sqb{\hbox{\hskip5pt\vrule width8pt height6pt depth1.5pt \hskip1pt}}
\def\cqfd{\ifmmode\hbox{\hfill\sqb}\else{\ifhmode\unskip\fi%\nobreak\hfil
\penalty50\hskip1em\null\nobreak\hfil\sqb\parfillskip=0pt\finalhyphendemerits=0\endgraf}\fi}

\newtheorem{claim}{Claim}

\usepackage[utf8]{inputenc}

\begin{document}
\title{Protected load-balancing problem: Neural-network based approximation for non-convex optimization} 
 
\author{
\IEEEauthorblockN{Youcef Magnouche, Sébastien Martin, Jérémie Leguay }
\IEEEauthorblockA{Huawei Technologies Ltd., Paris Research Center, France}}

\maketitle

\begin{abstract}
Nowadays, centralized Path Computation Elements (PCE) integrate control plane algorithms to optimize routing and load-balancing continuously. 
When a link fails, the traffic load is automatically transferred to the remaining paths according to the configuration of load-balancers. In this context, we propose a load-balancing method that anticipates load transfers to ensure the protection of traffic against any Shared-Risk-Link-Group (SRLG) failure. The main objective of this approach is to make better use of bandwidth compared to existing methods. It consists in reserving a minimum amount of extra bandwidth on links so that the rerouting of traffic is guaranteed. We propose a non-linear non-convex model for the problem of minimizing the bandwidth reservation cost. We introduce a new approximation approach based on a neural network to convexify the problem and apply Kelley’s cutting plane method to solve the problem. Finally, we show that our algorithm significantly improves the CPU time against a compact model solved using the SCIP solver.
\end{abstract}


\maketitle  
\thispagestyle{plain}
\pagestyle{plain}

\section{Introduction}
Load-balancing plays a crucial role in improving the utilization of telecommunication networks. It basically consists in splitting traffic over multiple paths to make a better use of network capacity. In modern  network architectures, Software-Defined Networking (SDN) controllers or Path Computation Elements (PCE)~\citep{paolucci2013survey} integrate control plane algorithms to optimize routing and load-balancing. These centralized control  entities acquire, thanks to network monitoring protocols, a global view of the network to decide whether it is necessary to split traffic and the most efficient way to do it. 

Typically, load-balancing is implemented inside network devices, such as switches and routers, using two techniques, hash-based splitting for Equal or Unequal Cost Multi-Pathing (ECMP or UCMP)~\citep{medagliani2016global} or Weighted Cost Multi-Pathing (WCMP)~\citep{zhou2014wcmp}. In hash-based splitting, a hash is calculated over significant fields of packet headers and used to select outgoing paths. Multiple forwarding rules, also called \textit{buckets}, can be configured for each path so as to customize split ratios. In weighted cost multi-pathing, load-balancing weights are used to define what portion of traffic must be sent over each path. In this case, more sophisticated mechanisms are required in the data plane to follow the utilization of paths and adjust decisions. For elastic flows, in general, once a decision is taken for a flow, all packets follow the same decision (same path) to avoid packet re-ordering issues. For real-time traffic, packet-level load balancing may be applied and controlled through an entropy~\cite{rfc6790} field in packet headers for hash-based splitting.

In practice, when a failure happens, the split ratios of an affected tunnel (or any traffic aggregate) are automatically adjusted, and the load is transferred to the set of surviving paths according to the load balancing configuration before failure (e.g., load-balancing weights or buckets). If failure scenarios are not anticipated well, the corresponding load transfers may generate congestion and induce performance degradation (e.g., packet loss).
To prevent this and protect traffic against failures, classical mechanisms~\citep{NARAGHIPOUR20082360} for single path routing can be applied in the context of load balancing. In this case, each path is protected by one or several backup paths. A typical example is 1+1 protection~\citep{zhou2000survivability} where traffic is sent over two paths with full duplication or a specific encoding~\cite{kamal2010overlay}. However, these solutions are not tailored to load balancing. Taking a conservative approach, they consider each path as an individual tunnel that needs to be protected.  Therefore, they can lead to a high bandwidth utilization or to the inability to protect  traffic due to the lack of capacity. 

To improve bandwidth utilization, we propose a new protection mechanism for load balancing, that makes an efficient use of bandwidth. The general idea is to configure, for a set of tunnels, “safe” split ratios to avoid traffic loss in case of  failure scenarios defined as Shared Risk Link Groups (SRLG). Each failure scenario consists in a set of links sharing critical resources (e.g., physical fiber) that could all fail together in case of outage. 
The objective of our mechanism is to globally optimize routing paths and
split ratios by anticipating all the possible load transfers that could happen in case of failures. %We propose two variants of the solution where the reserved bandwidth can be shared or not among failure scenarios. 
The main benefits of this approach are the following: 1) no distinction between primary and backup paths is needed anymore, 2) it can work with any load balancing solution as long as split ratios and routing paths can be controlled, and 3) strict protection, if needed, can be achieved by reserving protection bandwidth on each path.

In the rest of this paper, we introduce a new  protection mechanism for load-balancing and study the associated optimization problem that a network controller has to solve to compute routing paths and split ratios. For a set of tunnels, it  ensures that the utilized (or reserved) bandwidth is of minimum cost while protecting traffic against a set of failure scenarios. Overall, we present the following contributions: 
\begin{itemize}    
     \item First, we introduce a new protected load balancing mechanism for multiple tunnels where reservations can be shared across tunnels.
    \item Then, we formulate the  associated optimization problem to minimize the total bandwidth reservation cost using non-linear and non-convex programming.
    \item To efficiently solve the problem, we approximate the non-convex constraints by convex inequalities using a neural network, and we apply Kelley's cutting plane method \citep{kelley1960cutting}.
    \item Finally, we present computational results on realistic instances that indicate that the proposed algorithm finds good solutions while improving significantly the CPU time, compared to the compact model solved using SCIP~\citep{GamrathEtal2020OO}, and the quality of the solutions compared to a linear approximation of the problem.
\end{itemize}

The paper is organized as follows. Sec.~\ref{related} presents the related work. 
Sec~\ref{sec:shared} introduces the protected load-balancing mechanism. Sec~\ref{sec:optimization} gives some definitions and presents the associated mathematical model for the protected load-balancing problem.  Sec.~\ref{sec:approximation_non_convex} describes the use of a neural network for approximating the non-convex constraints by convex inequalities. Sec.~\ref{sec:neural_kelley_cutting_plane} describes the designed algorithm based on Kelley's method for convex problems. 
Sec.~\ref{sec:numerical-results} reports numerical results and Sec.~\ref{sec:conclusion} concludes the paper. 

\section{Related Work}
\label{related}

A natural way to protect load-balancing paths is to consider a backup path for each primary path involved in the load-balancing of each tunnel. However, this mechanism is equivalent to 1+1 in terms of bandwidth utilization. To further optimize resources, 1:1 protection reserves bandwidth for backup but shares it among tunnels. Upon detection of a failure, a notification towards the head end router of tunnels triggers the re-routing of the protected traffic on backup paths.
A Shared-Backup protection (SBP) mechanism allows tunnels with disjoint working paths to share the same bandwidth reservations for their backup paths~\citep{mello2005dynamic} if they are not used simultaneously in case of failure. In~\citep{doucette2005advances} authors propose an Integer Linear Program (ILP) model for the SBP problem to select a backup and a working path for every tunnel such that the cost of bandwidth reservations is minimized. This approach can be extended to $m$:$n$  protection, where $n$ recovery paths can be shared to protect $m$ working paths. However, to our knowledge, no bandwidth efficient method has been proposed to protect tunnels that are load-balanced over multiple paths, beyond the protection of individual paths, considered as independent tunnels.

\begin{figure}[t]
\centering
\includegraphics[scale=0.35]{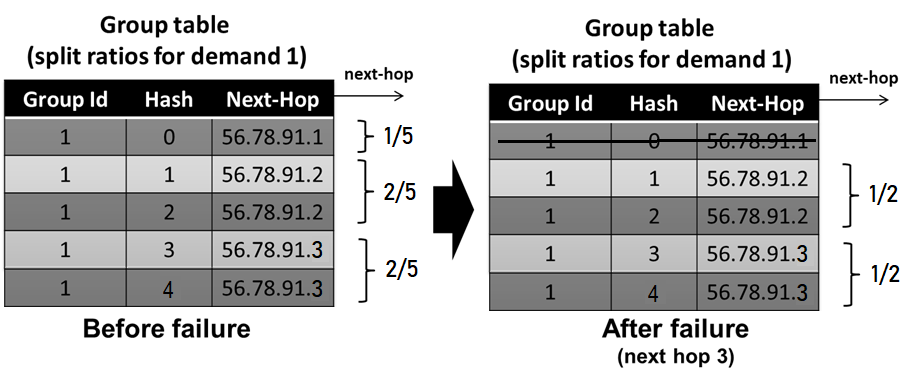} 
\caption{Impact of a failure on split ratios in UCMP.}\label{hash-table-failure}
\end{figure} 

Control plane algorithms have been proposed~\citep{medagliani2016global, kang2015efficient} to optimize the configuration of split ratios, but they do not anticipate all possible failure scenarios, and the associated load transfers among paths, when making decisions. Some works~\citep{kang2014niagara} have proposed to gracefully update split ratios after each failure, but they consist in reactive approaches that can be slow to restore services.

\section{Protected load balancing}
\label{sec:shared}

This section provides a motivation example and illustrates the main idea of our protected load balancing mechanism.

\subsection{Hash-based splitting}

Without loss of generality, we provide an example for hash-based splitting with uneven load balancing, i.e. UCMP. However, the same applies to weighted cost multi-pathing. Fig.~\ref{hash-table-failure} presents an example in which load balancing is implemented in a so-called \textit{group table} inside the data plane of network devices. Each tunnel that needs to be load-balanced over multiple paths is associated to a group with multiple entries in this table. Each entry determines the next-hop of a flow as a function of a hash value calculated over its packet headers. An entry can be duplicated several times. The ratio between the number of rules for a next-hop and the total number of rules for the group determines the split ratio.  
As we can see, when the link associated to the next-hop $56.78.91.1$ fails, the load for demand $1$ (or tunnel $1$) associated to the group $1$ is transferred to the remaining next-hops, modifying their split ratios from ($\frac{1}{5}$, $\frac{2}{5}$, $\frac{2}{5}$) to ($0$, $\frac{1}{2}$, $\frac{1}{2}$).

\subsection{Protection mechanism}

\begin{figure}[t]
	\centering
	\includegraphics[scale=0.3]{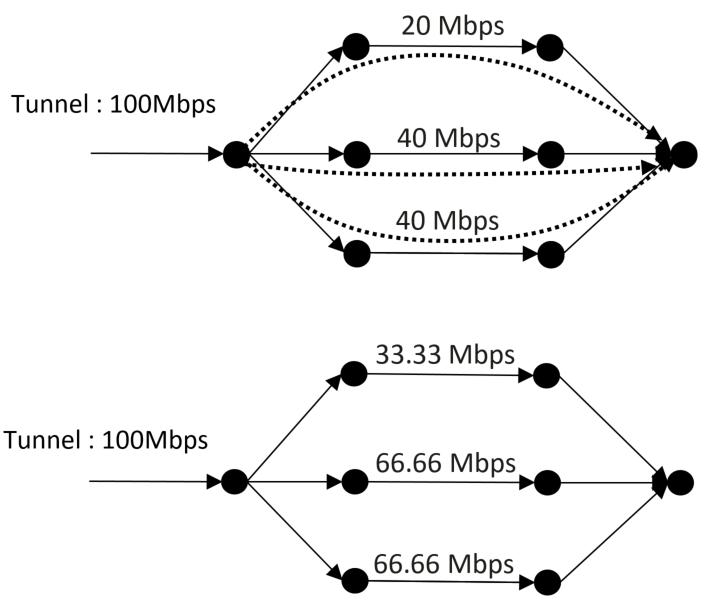} 
	\caption{Example of bandwidth reservations (bottom figure) for a given set of split ratios (top figure), for a tunnel load balanced over $3$ paths and protected against $1$ link failure.  }\label{bandwidth-reservation}
\end{figure}  

In practice, congestion can happen after each failure and traffic may be lost. For this reason, we propose to calculate split ratios so as to ensure that traffic will not face  congestion after every possible failure scenario. To do this, we need to calculate how much bandwidth will be utilized over each path in the worst case. This bandwidth can be reserved, when MPLS is used, or simply considered in the calculation of split ratios for a more best effort implementation.
Fig.~\ref{bandwidth-reservation} illustrates the computation of  bandwidth reservations for a protection against $q$-link failures (with $q\in \mathbb N$, see~\citep{yang2014keep}). In this example, we consider one tunnel load balanced over $3$ disjoint paths and $q=1$.  
At the top, we consider a given set of split ratios for a tunnel over $3$ paths. At the bottom, we display the \textit{bandwidth reservations} to cover all possible link failures. Note that, as all paths are disjoints, all links of a path require the same bandwidth reservation.  
Since the splits are unequal, when a link fails, new splits ratios are computed on the remaining paths based on the initial splits ratios. 
Let $p_1, p_2$ and $p_3$ be the top, middle and bottom paths in the example.
Since there are only $3$ paths, bandwidth reservations can be computed for each path by considering only two scenarios (e.g., for $p_1$ we consider two scenarios, when $p_2$ or $p_3$ fails). Formally, the bandwidth reservation on every link of path $p_1$ can be computed using the following formula:

{
\scriptsize{
$$
 (\text{Traffic on }p_1) +
 $$\\
 \vspace{-0.5cm}
 $$
max\{
 \frac{(\text{Traffic on }p_2)\times(\text{Traffic on }p_1)}{(\text{Traffic on }p_1 + \text{Traffic on }p_3)}
,
 \frac{(\text{Traffic on }p_3)\times(\text{Traffic on }p_1)}{(\text{Traffic on }p_1 + \text{Traffic on }p_2)}
\}\label{reservation_computation}
$$
}
}
The bandwidth reservation on every link of path $p_1$ is given as follows: 
$$ 20 + max\{ \frac{40 \times20}{20 + 40}, \frac{40\times 20}{20+40} \} = 33.33 \text{ Mbps}$$ 
Similarly for the two other paths, the bandwidth reservation on every link of path $p_2$ and $p_3$ is $66.66$ Mbps, when considering all possible link failures. 

In the example, we can observe that this protection mechanism is more efficient than 1+1 as it requires only $33.33 \times 3 + 66.66 \times 6 = 499.95$ Mbps of bandwidth over all paths instead of $100 \times 6 = 600$ Mbps.
Moreover, we observe that bandwidth reservations depend on the initial split of traffic. Hence, different split ratios may increase or decrease the total reservation cost. In the worst case, the bandwidth reservation cost cannot be larger than that of 1+1. Furthermore, the gain against 1+1 does not  depend only on the load-balancing. Indeed, for $q$-link protection, at least $q + 1$ disjoint paths (without common links) are required to ensure the protection. While it requires at least $q+2$ disjoint paths to hope obtaining a lower bandwidth reservation than 1+1. Indeed, if there are $q+1$ paths and $q$ links fail, they may destroy $q$ paths for a tunnel, and most of the traffic must be rerouted on the remaining path. Therefore, the required bandwidth reservation is $200\%$, as for 1+1. 

In the rest of the paper, we will focus on the optimization of bandwidth reservations for a set of tunnels with protection against 
failure scenarios defined as Shared Risk Link Groups (SRLG). 
When several tunnels are considered in the same network, it may happen that the SRLG failures do not impact all tunnels together. Therefore, under some conditions, the same reserved bandwidth can be used by two different tunnels, leading to a bandwidth reservation saving. This mechanism is called \textit{shared protection}. The same approach can be considered for the \textit{unshared} case.  

\section{Optimization problem} \label{sec:optimization}

We first give some definitions and notations used throughout this paper. We then formulate the optimization problem to calculate "safe" split ratios.

\subsection{Definitions \& notations} 
We consider a network represented by a simple graph $G=(V, E)$, where $V$ is the set of nodes and $E$ the set of links, and a set of tunnels $K$. Let $K^+$ (resp. $K^-$) be the subset of protected (resp. unprotected) tunnels in $K$ such that $K=K^+ \cup K^-$. Every tunnel $k\in K$ is defined by a source $s_k$, a destination $t_k$, a traffic demand $d_k\in \mathbb R^+$ and a set of paths $P^k$ between $s_k$ and $t_k$ (not necessarily disjoints). For an edge $e\in E$, let $b_e\in \mathbb R^+$ be the associated bandwidth capacity and $c_e\in \mathbb R^+$ be the unit cost for the reserved bandwidth. For $p\in P^k$, let $c_p\in \mathbb R^+$ be the routing cost associated with $p$, for instance, the IGP cost. % 
Let denote by $S\subseteq E$ an SRLG (a set of links that can fail together) and by $\mathcal S$ the set of all given SRLGs against which we need to protect traffic. 
Let $P^k_{S}\subseteq P^k$ be the subset of paths of tunnel $k\in K$ intersecting SRLG $S\in \mathcal S$. %
Let $P^k_e\subseteq P^k$ be the subset of paths of tunnel $k\in K$ containing $e\in E$. If an SRLG \textit{fails}, all its links fail. A path \textit{fails}, if at least one of its links fails. An amount of traffic is said to be \textit{rejected} if it passes through a failed path (after failure and transfer of the load). Similarly, traffic is called \textit{accepted} if it does not cross any failed path. Two paths are called \textit{disjoint} if they do not intersect a common SRLG.

\subsection{Problem formulation}

The protected load-balancing problem consists in determining the split ratios for every tunnel $k\in K$ over all paths $P^k$ such that the total bandwidth reservation cost of the protected tunnels and the routing cost is minimum. When multiple tunnels cannot be affected by an SRLG failure simultaneously (thanks to disjointness of the paths),  backup resources can be shared among these tunnels, which decreases the total bandwidth reservation. 

For example, let consider two tunnels $k_1$ and $k_2$ with path sets $\{p_1^1, p^1_2\}$ and $\{p^2_1, p^2_2\}$, respectively, such that $p_1^1$ and $p^2_1$ are totally disjoints, and $p_2^1$ and $p^2_2$ share a common link $e\in E$. Clearly, for $1$-link protection, the paths $p_1^1$ and $p^2_1$ cannot fail simultaneously. Therefore, a common bandwidth reservation on $e$ can be used by tunnel $k_1$ when $p_1^1$ fails, and by tunnel $k_2$ when $p_1^2$ fails.

The shared  reservations must be computed for each link $e\in E$. Let denote by $w^{ek}_S\in [0,1]$ the ratio of $d_k$ passing through $e$ but not crossing $S$, before $S$ fails. Also, let denote by $w^k_S\in [0,1]$ the ratio of $d_k$ passing through SRLG $S$. 
After $S\in \mathcal S$ fails, the new amount of traffic passing through $e$ is the sum of all rerouted traffic on $e$ of all tunnels $K^+$.
Formally, the new amount of traffic on $e$ after $S$ fails is equal to $\sum\limits _ {  \forall k\in K^{+}}  (  d_k w^{ek}_{S} +  d_kw^k_{S}\frac{  w^{ek}_{S}}{  1 - w^k_{S} })$.
The bandwidth reservation on link $e$ for tunnel $k$ is equal to the maximum amount of traffic after every SRLG failure. 
 The bandwidth reservation $w_e$ on the link $e\in E$ is equal to 
$\max \limits _{S \in \mathcal S}\{ \sum\limits _ {  \forall k\in K^{+}}  \{  d_k w^{ek}_{S} +  d_kw^k_{S}\frac{  w^{ek}_{S}}{  1 - w^k_{S} }  \}\} $. Since the bandwidth reservation cost is minimized, the equality can be transformed to a set of inequalities $\sum\limits_ { \forall k\in K^{+}} (d_k w^{ek}_{S} + d_k w^{ek}_{S} \frac{ w^k_{S} }{  1 - w^k_{S}  }) \leq w _e$ for all $e\in E$ and for all $S\in \mathcal S$. This is equivalent to $\sum\limits _ {  \forall k\in K^{+}} \frac{ d_k w^{ek}_{S}  }{  1 - w^k_{S}  }\leq w _e$ for all $e\in E$ for all $S\in \mathcal S$.\\

In the following, we present the mathematical model for the protected load-balancing problem to compute 1) split ratios for every path of every tunnel and 2) shared bandwidth reservations for every link. %Consider the following variables. 
Let $x_p^k \in [0,1]$ be the split ratio of path $p$ used by tunnel $k\in K$ and $w_e\in \mathbb R$ be the shared bandwidth reservation on link $e\in E$. %$w^{ek}_{S}\in \mathbb R^+$ be the ratio of $d_k$ passing through $e\in E$ before $S\in \mathcal S$ fails, for tunnel $k\in K^+$; $w^k_{S}\in \mathbb R^+$ be the ratio of $d_k$ passing through SRLG $S\in \mathcal S$ for tunnel $k\in K$. 
The objective is to minimize the cost of the reserved bandwidth for the protected tunnels, together with the total paths cost.
The problem is equivalent to the following non-linear program:

{
\small
{
\begin{align}
&\min \sum_{e\in  E} c_e w_e   +\sum_{k\in K^{+}\cup  K^{-}}d_k\sum_{p\in P^k} c_p x_p^k & & \nonumber\\   
\label{1}&  \sum_{p\in P^k} x_p^k = 1  &&  \forall k\in K^{+}\cup  K^{-},\\ 
\label{2}&\sum_{p\in P^k_{S}} x^k_p  \leq w^k_{S}&&  \forall S\in \mathcal S, k\in K^{+},\\
\nonumber&  \sum_{p\in P^k_{e}\setminus P^k_{S}} x^k_p \leq w^{ek}_{S}&& \\
\label{3}& \qquad\qquad\qquad \forall e\in E, k\in K^{+}, \forall S\in \mathcal S ,  &&\\
\label{4}& \sum_{k\in K^{-}}d_k\hspace{-2mm}\sum_{p\in P^k_{e} \setminus P^k_{S}} \hspace{-2mm} x^k_p + \sum_{k\in  K^{+}} \frac{  d_k w^{ek}_{S}  }{  1 - w^k_{S}} \leq b_e  && \forall e\in E , S\in \mathcal S ,\\
\label{5}& \sum_{k\in  K^{+}} \frac{  d_k w^{ek}_{S}  }{  1 - w^k_{S}}  \leq w_e  &&  \forall e\in E , S\in \mathcal S, \\
\label{6}& w^k_{S} \leq 1 - \epsilon  && \forall k\in K^{+} , S\in \mathcal S.
\end{align}   
}
}
Equalities \eqref{1} ensure that all traffic is split on the paths of each tunnel. Inequalities \eqref{2} compute the ratio of $d_k$ passing through $S\in \mathcal S$ for tunnel $k\in K^+$. Inequalities \eqref{3} compute the ratio of $d_k$ passing through $e\in E$ after $S\in \mathcal S$ fails, for tunnel $k\in K^+$.
Inequalities \eqref{4} represent the link capacity constraints. 
Inequalities \eqref{5} compute the bandwidth reservation of the protected tunnels on every link.
Inequalities \eqref{6} ensure that every tunnel does not send all its traffic through the same SRLG, where $0 \leq \epsilon < 1$ is small enough.
 
The above mathematical model in non-convex and non-linear program. This is due to Constraints \eqref{4} and \eqref{5}. When $|K^+| = 1$ and $|K^-| = 0$, the constraints are of the form $f(x)\leq t$ where $f(x, y)=\frac{x}{1-y}$ such that $x\in [0,1]$ and $y\in [0, 1[$. Consequently, Constraints \eqref{4} and \eqref{5} are non-convex.

\section{Approximation of non-convex constraints}\label{sec:approximation_non_convex}

A non-linear programming model refers to a mathematical program with at least one non-linear function \citep{griva2009linear}. 
Non-convex programming represents one of the most challenging fields of optimization \citep{liberti2008introduction} and no efficient approach is available to derive the global optimum \citep{lin2013review}. The best-known method for this type of problem is the Spatial Branch-and-Bound. It consists in dividing the program into several convex approximations within a branch-and-bound tree. 
For decades, researchers are studying the approximation of non-linear functions \citep{devore1998nonlinear}. The goal is to replace a difficult function by another one having a much lower complexity. Several methods have been proposed in the literature such as  the approximation with piecewise constants, wavelets, linear and non-linear piecewise approximation, etc.

 %%%%%%%%%%%%% NEURAL NETWORK FOR CONVEXIYING %%%%%%%%%%
A well-known theorem in this field is the universal approximation theorem \citep{csaji2001approximation}. It states that an arbitrary continuous function, defined on $[0,1]$ can be arbitrary well uniformly
approximated by a multilayer feed-forward neural network with one hidden layer
(that contains only a finite number of neurons) using neurons with arbitrary activation
functions in the hidden layer and a linear neuron in the output layer. Therefore, in the rest of this section, we investigate the approximation of our non-convex non-linear function $f(x, y)=\frac{x}{1-y}$ using an artificial neural network to obtain a convex function. This will allow us to convert the mathematical model given in Sec.~\ref{sec:shared} into a convex non-linear model that can be solved optimally using Kelley's algorithm (see Sec.~\ref{sec:neural_kelley_cutting_plane}). To the best of our knowledge, there is no such work in the literature for solving non-convex non-linear problems that combines Kelley's algorithm and a neural network based approximation.

\subsection{Neural network based approximation}

Artificial neural networks (ANNs) are composed of artificial neurons which are conceptually derived from biological neurons. Each artificial neuron has inputs and produces a single output which can be sent to multiple other neurons. 

The input of each neuron is passed through an  activation function  (usually non-linear) to produce the output. Several well-known activation functions can be used such as Relu, Identity, arcTan, PRelu, SoftPlus, SoftMax, etc. The network consists of connections between neurons where 
a weight is assigned to represent its relative importance. A given neuron may have multiple input and output connections.

\begin{figure}[t]
    \centering
    \includegraphics[scale=0.35]{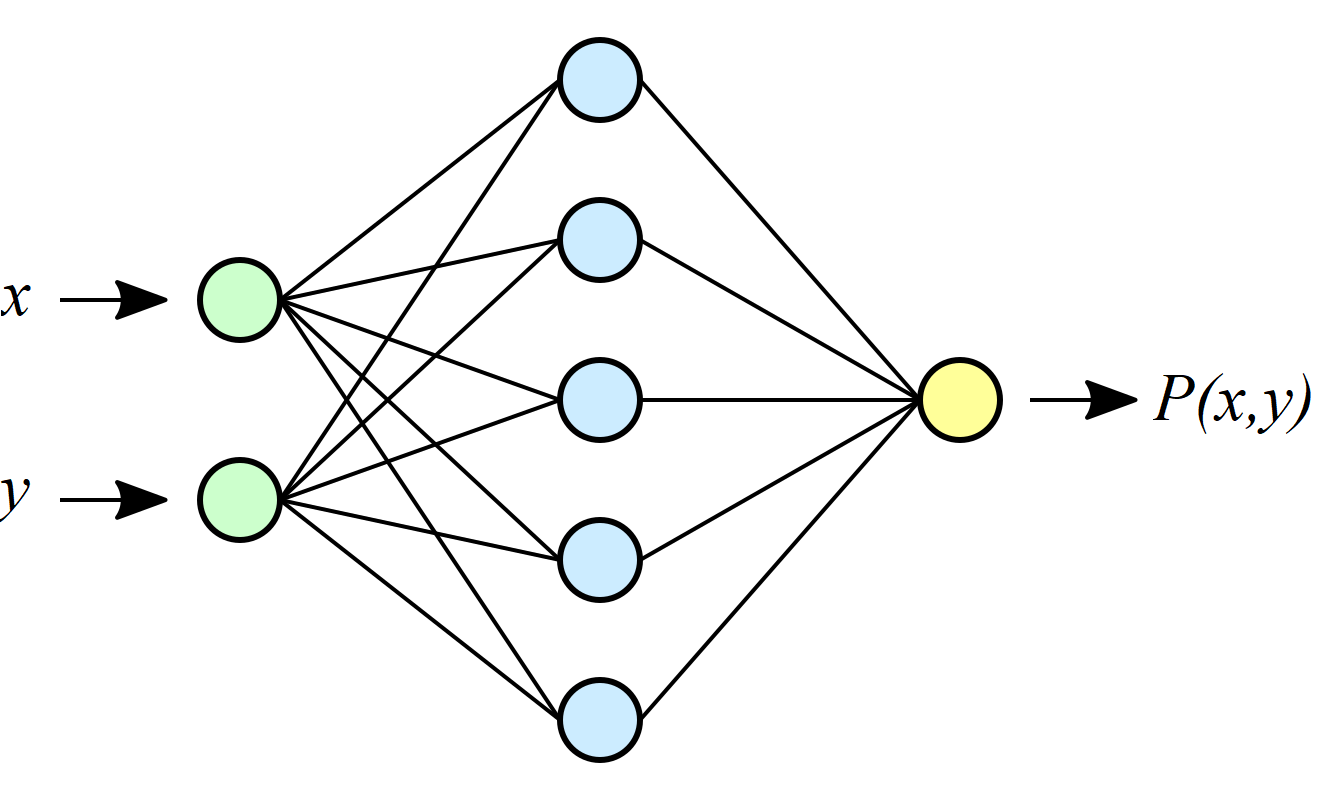}
    \caption{Feed-forward neural network  with 1 hidden layer.}
    \label{simple_neural_network}
\end{figure}

A bias term can be added to the input of each neuron. Fig.~\ref{simple_neural_network} illustrates a neural network with 8 neurons. 
The green, blue and yellow neurons represent, the input, hidden and output layers, respectively. In the above example, there is only 1 hidden layer of 5 neurons. \\
Let $x, y\in \mathbb R$ be the two input values of the ANN, the estimated function $P(x,y)$ is given as follows: 
$$P(x,y) = \sum \limits _{i=1}^{n} a_i f_i(a_x^ix + a_y^i y + b_i) + b$$
where $a_x^i$ represents the weight between input $x$ and $i^{th}$ neuron of the hidden layer; $a_y^i$ represents the weight between input $y$ and $i^{th}$ neuron of the hidden layer; $b_i$ represents the bias of the $i^{th}$ neuron of the hidden layer; $a_i$ represents the weight between the output of $i^{th}$ neuron of the hidden layer and the output neuron; and $b$ : represents the bias of the output neuron.

\begin{claim} \label{claim:convexity}
If $a_i\geq 0$ and $f_i$ is convex for all $i\in \{1,\dots, n\}$, then $P(x,y)$ is also convex. 
\end{claim}
 
In the following, we approximate the non-convex non-linear function $f(x,y) = \frac{x}{1-y}$ by a convex non-linear function using a simple feed-forward neural network. \\

For our numerical results, we used the following artificial neural network configuration 
\begin{enumerate}[1-]
    \item Number hidden layers: we use only 1 hidden layer.
    \item Activation functions: we considered the following types of convex functions on $\mathbb R$ : $x^{2i}$ for all $i\in \{0, \dots, 10\}$, $e^x$, $Relu(x) = \max\{x, 0\}$.
    \item Number of neurons in the hidden layers: after several attempts, we used $5$ activation functions of each type. 
    \item Loss function: we  used the mean squared function, which is $(y_{true} - y_{pred})^2$, as commonly used for regressions.
    \item Labels: we generate the data by varying the values of $x$ and $y$ in the associated domains which provides the features. The labels are obtained using the original function $\frac{x}{1-y}$ for every feature $(x,y)$. Precisely, we consider a set $S_x$ (resp. $S_y$) of 100 evenly spaced numbers over the interval $[0.05, 1.0]$ (resp. $[0.00, 0.99]$). The features represent all pairs $(x',y')$ such that $x'\in S_x$, $y'\in S_y$ and $ x' + y' \leq 1$. 
    %Note that when $x\leq 0.05$, the approximation is very poor. 
    \item Optimizer: we used Adam (Adaptive Moment Estimation Algorithm).
    \item Positive weights: to preserve the convexity of the estimated function, weights $a_i$ for $i=\overline{1,n}$ must be positive. 
    \item Number of iterations: we used 300.
\end{enumerate}

\begin{figure}[t]
\centering    
\includegraphics[scale=0.1]{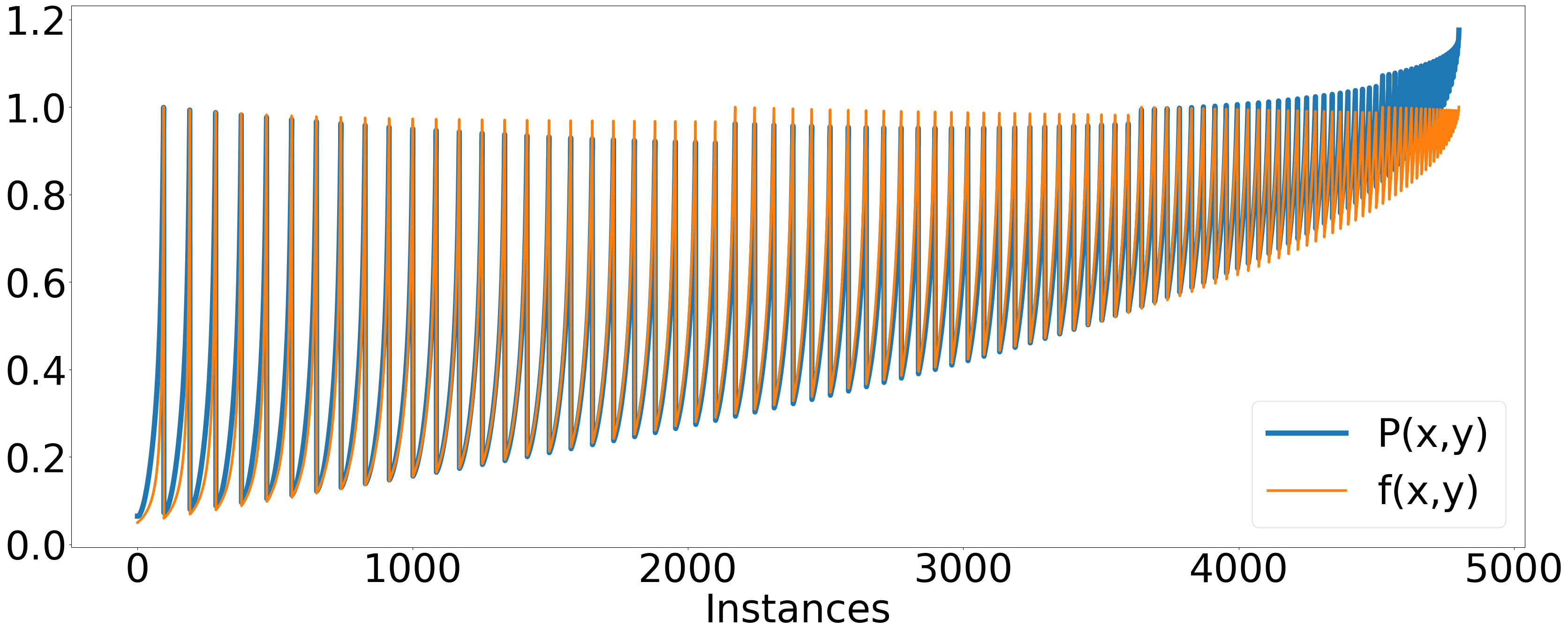}  
\caption{Comparison between the original function $f(x,y)$ and the estimated one $P(x,y)$ obtained from the neural network.}
\label{ANN}
\end{figure}

Fig.~\ref{ANN} displays two curves representing the function values given by the original and the approximation ones, over around 5000 points. We can see that the approximation function fits perfectly $\frac{x}{1-y}$ on all the points. However, when $x$ is near to $0.05$ and when $y$ is near to $0.00$, the gap increases a bit to reach $6\%$. 

As one can notice, the approximation function we found is near or above $f(x,y)$. This is important to ensure the feasibility of  configurations for  protected load-balancing. In our case, this means that the bandwidth reservations are slightly over-approximated (conservative). If the approximation were below $f(x,y)$ we could define a custom loss function to force the approximation to be always (almost) above $f(x,y)$.\\

\begin{figure}[t]
\centering
\includegraphics[scale=0.08 ]{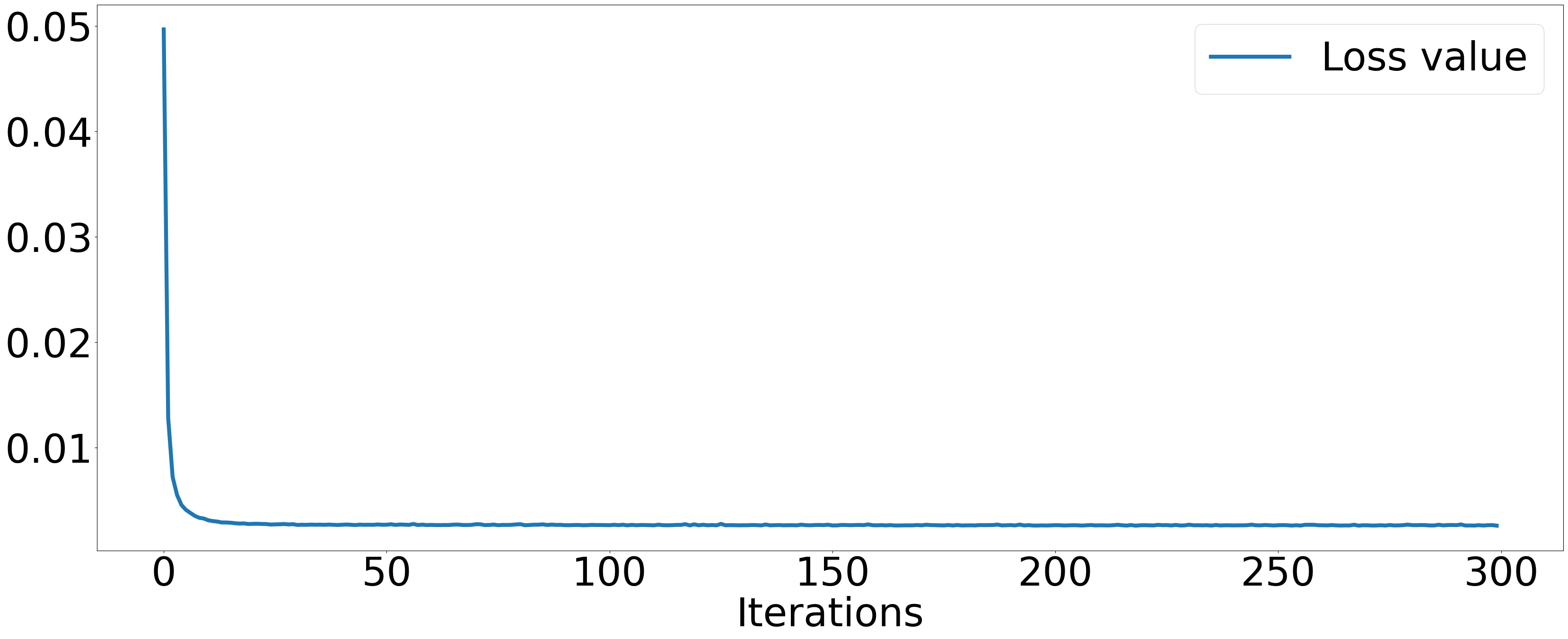}
\caption{Evolution of the loss function during the training phase.}
\label{ANN:loss}
\end{figure}

Fig.~\ref{ANN:loss} shows the evolution of the loss value over the iterations in the training phase. Clearly, the loss value converges to $0$ after $20$ iterations. 

\subsection{Convex transformation for protected load balancing}

The learned approximation function is as follows\\

\begin{footnotesize} 
 $P(x, y) =  0.20898512 (0.84153354 x -1.2163565 x -0.1172726) + 0.11398053 (-0.54810554 x -0.73203504 x -0.111881435)^2 + 0.22500123 (-0.32912576 x -0.4372456 x -0.10842324)^2 + 0.31306696 (-0.5892102 x -0.78629917 x -0.13606171)^2 + 0.12807561 (-0.7460143 x -0.99790335 x -0.06382404)^2 + 0.13445288 (0.32067707 x + 0.42634684 x -0.053649783)^2 + 0.08919548 (-0.008583701 x -0.014321463 x + 0.011760315)^4 + 0.0005710255 (-0.16931695 x -0.20765431 x + 0.13124172)^4 + 0.0006219578 (-0.19715643 x -0.25719136 x + 0.16870873)^4 + 0.110854276 (0.08626526 x + 0.11822398 x + 0.0036681416)^6 + 0.18170816 (-0.1423485 x -0.16759561 x + 0.08233984)^6 + 0.21071501 (-0.27259338 x -0.33094826 x -0.039481297)^8 + 0.017079473 (0.6349315 x + 0.7413781 x -0.18566795)^{10} + 0.04180262 (0.36131313 x + 0.4217615 x -0.111623645)^{10} + 0.023078842 (0.12757984 x + 0.080746045 x )^{12} + 0.067266844 (-0.29820144 x + 0.051307905 x + 0.08951803)^{12} + 0.24862847 (0.49615633 x + 0.5693221 x -0.08240397)^{14} + 0.10745613 (-0.68662584 x -0.7797363 x + 0.11783628)^{16} + 0.22991414 (0.885376 x + 1.0115623 x + 0.016353546)^{18} + 0.20432162 (-0.08416062 x + 0.11819947 x )^{18} + 0.19670802 (0.52702695 x + 0.59805554 x -0.048327893)^{18} + 0.13163748 (-0.7192018 x -0.8070603 x + 0.14870796)^{18} + 0.11298036 (-0.59983516 x -0.6753686 x + 0.050704874)^{20} + 0.025386946 e^{-0.4838321 x -0.6530228 x -0.6168387} + 0.008721772 e^{-2.1046681 x -2.9321532 x -4.5336885} + 0.010362742 e^{-2.1843915 x -3.0633261 x -4.80528} + 0.002250989 e^{-1.0597663 x -1.3402514 x -1.7501163} + 0.14911664 e^{-0.028126838 x -0.044313956 x -0.24021097} + 0.15225288 \max\{0.0, 0.5040369 x + 0.4609523 x -0.04698109\} + 0.7161469 \max\{0.0, 0.40694946 x + 0.5698667 x -0.37897623\} + 0.1736147 \max\{0.0, 0.62029403 x + 0.42131162 x -0.050037168\} + 0.6821202 \max\{0.0, 0.64257723 x + 0.026824359 x + 0.0063459207\} -0.087192556
$ 
\end{footnotesize}  

Consequently, for the protected load balancing problem, Constraints \eqref{4} and \eqref{5} are approximated by the following constraints, for all $S\in \mathcal S$ and $e\in E$:

\begin{align}
\label{approx:4}& \sum_{k\in K^{-}}d_k\sum_{p\in P^k_{e} \setminus P^k_{S}} x^k_p + \sum_{k\in  K^{+}}  d_kP(w^{ek}_{S},w^k_{S}) \leq b_e,\\
\label{approx:5}& \sum_{k\in  K^{+}}  d_kP(w^{ek}_{S},w^k_{S})  \leq w_e.
\end{align}

By replacing the non-convex functions with convex functions, we now are able to solve the protected load-balancing problem using a convex optimization method.

\section{Kelley's Cutting plane algorithm (NKCP)}\label{sec:neural_kelley_cutting_plane} 

 %%%%%%%%%%%%% CONVEX Constrained OPTIMIZATION %%%%%%%%%%%%%%%%%%%%%
Efficient methods have been proposed for convex non-linear programs, those composed of a concave objective function (in case of maximization problems), or convex objective function (in case of minimization problems) and a set of convex constraints. The most important one are Bundle \citep{bertsekas1997nonlinear}, Subgradient projection \citep{kiwiel2007lagrangian}, Interior-point \citep{nesterov1994interior}, Outer-Approximation \citep{duran1986outer} and Ellipsoid \citep{grotschel1981ellipsoid} methods. In \citep{kelley1960cutting}, the author suggests an optimization method, called Kelley's cutting plane method, to solve optimally the convex mixed integer non-linear problems. It consists in replacing the convex non-linear constraints by linear outer-approximations constraints to obtain a linear program. The outer-approximation function is a linear function that approximate a non-linear one at a particular point. Let $f(x)\leq 0 $ be a convex non-linear inequality, the associated outer-approximation inequality at point $x^*$ is 
$$  \triangledown f(x^*)^{\top} (x- x^*) + f(x^*) \leq  0 $$  

Consider the following model,
\begin{align}
\label{NLP:obj1}\min \qquad & cx   & & \\   
\label{NLP:1}&  Ax \geq b &\quad&  \\ 
\label{NLP:2}& f(x)\leq 0 &\quad&   \\
\label{NLP:3}& x\in \mathbb R.& &
\end{align}   
Constraints \eqref{NLP:1} represent the set of linear inequalities and  Constraints~\eqref{NLP:2} the set of convex non-linear constraints. 
The above model is equivalent to the following LP model, 
\begin{align}
\label{NLP:obj2}\min \qquad & cx   & &  \\   
\label{NLP:4}&  Ax \geq b &\quad&  \\ 
\label{NLP:5}& \triangledown f(x^*)^{\top} (x- x^*) + f(x^*) \leq 0 &\quad& \forall x^*\in X  \\
\label{NLP:6}& x\in \mathbb R.& &
\end{align} 
where $X=\{x\in \mathbb R | Ax\geq b\}$.

Linear program \eqref{NLP:obj2}-\eqref{NLP:6} has an exponential number of Constraints \eqref{NLP:5}. Kelley's method consists in generating Inequalities \eqref{NLP:5}, iteratively, thanks to the cutting plane algorithm. Note that Kelley's method is closely related to the outer-approximation method that requires, in contrast to Kelley's method,  solving non-linear sub-problems. Also, it is closely related to the extended cutting plane algorithm that has been proposed in \citep{westerlund1995extended} for MINLP. 
 %%%%%%%%%%%%% NON-CONVEX OPTIMIZATION %%%%%%%%%%%%%%%%%%
 The algorithm that we developed is for solving the non-linear non-convex model given in Sec.~\ref{sec:shared}. It is based on the Kelley's cutting plane algorithm, after the replacement of non-linear non-convex constraints by convex non-linear constraints (see, Sec. \ref{sec:approximation_non_convex}).

Once the convex non-linear program is obtained, Kelley's cutting plane algorithm will allow solving it using a linear solver. This latter involves new constraints called \textit{outer-approximation constraints}
that replace the convex non-linear constraints.  
However, they may be exponential in number which require a dynamic generation of these constraints in the model. Hence, we apply the cutting-plane algorithm to generate them iteratively. 
 First, we start with a linear model without any outer-approximation constraint, and then we solve the sub-problem (associated to the remaining linear program). Let $x(i)$ be the optimal solution of the sub-problem. The next step consists in finding the outer-approximation cut of \eqref{approx:4} and \eqref{approx:5} at point $x(i)$. If a  violated cut is found, it is added to the model and the procedure  repeats until no violated outer-approximation cut is found. 
 
Since Constraints \eqref{approx:4} and \eqref{approx:5} are the approximation of the associated original constraints, the total bandwidth reservation may be not accurate. Therefore, after the optimization phase, the exact bandwidth reservation is computed based on the split ratios of the obtained solution. 
 
Note that, the routing paths can be computed during the optimization thanks to the column generation~\cite{desaulniers2006column} algorithm. Indeed, the $x$ variables do not belong to \eqref{approx:4} and \eqref{approx:5} which allows preserving the structure of the pricing problem of the column generation algorithm. In this case, the global algorithm generates both cuts (outer-approximation) and columns (variables $x$). However, when producing numerical results for the next section we considered a set of pre-calculated paths for each tunnel.

%%%%%%%%%%%%%%%%%%%%%%%%%%%%%%%%%%%%%%%%%%%%%%%%%%%%%%%%%%%% 

\section{Numerical experiments}\label{sec:numerical-results}

In this section, we present the numerical results obtained from the compact non-linear non-convex models solved using the SCIP Optimization Suite 7.0 \citep{GamrathEtal2020OO} and the Non-convex Kelley's Cutting Plane (NKCP) algorithm introduced in Sec.~\ref{sec:neural_kelley_cutting_plane}. The linear programs in NKCP have been solved using Cplex {12.6} ~\citep{cplex}. The algorithms have been implemented in C++ and tested on a machine equipped with an Intel(R) Xeon(R) CPU E5-4627 v2 of $3.30$GHz and $504$GB RAM, running under Linux $64$ bits. We considered a time-limit of $10$ minutes and a maximum of  $1$  thread. Function approximation is done with Python 2.7.16 using Keras for the neural network optimization~\citep{chollet2015keras} and scikit-learn~\citep{scikit-learn} for the linear regression (used as a benchmark to highlight that a basic method to approximate the non-linear function may impact the algorithm's performance).

\subsection{Test instances}

To evaluate our algorithm, we use two types of instances: Internet Topology Zoo~\citep{zooTopology} and SNDLIB~\citep{SNDlib10}. The SNDLIB instances provide all traffic information (sources, destinations and traffic demands) with a number of tunnels between 10 and 462. However, for Internet Zoo topologies, all demands have been generated randomly (sources, destinations and traffic demands). We have varied a total of 10, 40 and 80 tunnels\footnote{All instances are available on the following public repository:\\
\url{https://github.com/MagYou/Protected-Load-Balancing.git}}.

While our implementation and models support protection against any set of SRLG failures, we considered $q$-link protection for the tests (resilience to any simultaneous $q\in \mathbb N$ link failures) and we varied the following parameters: 
 
\begin{itemize}
\item number of protected tunnels: $40\%$ and $80\%$ of the total,
\item paths $n \in \mathbb N$ per tunnel (if they exist): $3$ and $6$,
\item SRLGs: all combinations of $q$ links where $q$ is $1$ and $2$.
\end{itemize}

The paths are generated as follows. For every tunnel, $n'= 30 \times n $ shortest paths are first computed using Yen's algorithm, minimizing the number of hops (links). Then, we solve an optimization problem in order to maximize the disjointness of the paths, i.e, find $n$ paths minimizing the maximum number of paths sharing a common SRLG.

\subsection{Test results}

\begin{figure}[t]
    \centering 
    \includegraphics[scale=0.23]{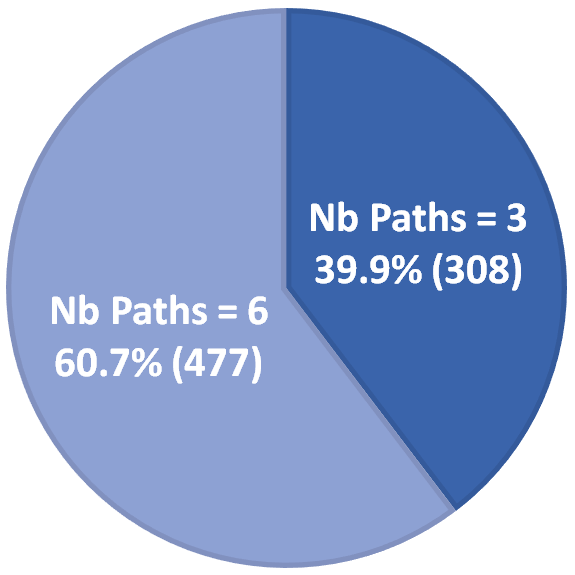} ~~~~
    \includegraphics[scale=0.23]{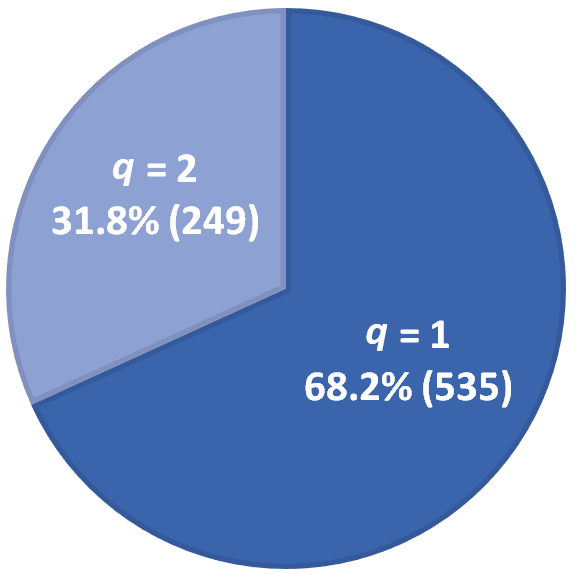}
    \caption{Number of instances solved using NKCP algorithm.}
    \label{fig:pie_nkcp}
\end{figure}

\begin{figure}[t]
    \centering 
    \includegraphics[scale=0.23]{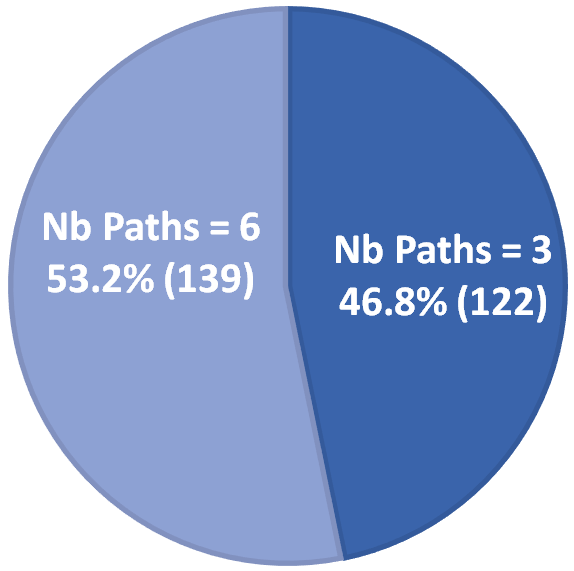} ~~~~
    \includegraphics[scale=0.23]{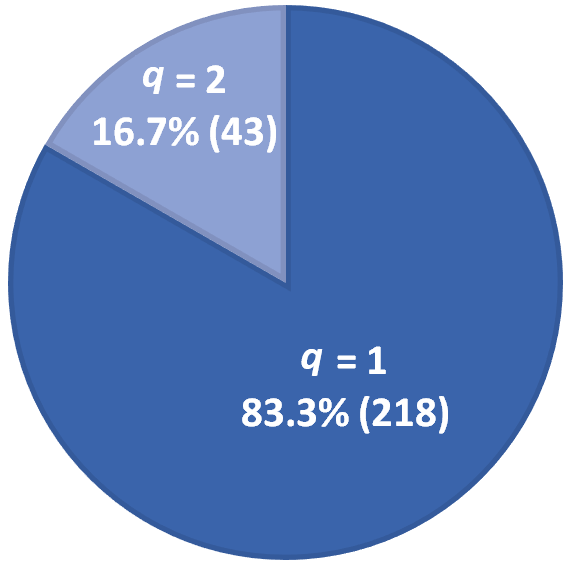}
    \caption{Number of instances solved using SCIP.}
    \label{fig:pie_scip}
\end{figure}

In the following, we compare the performance of the NKCP algorithm against the SCIP solver (compact model).
In all the following plots, we compare the instances solved by either NKCP or SCIP. Fig.~\ref{fig:pie_nkcp} and \ref{fig:pie_scip} display the number of solved instances for the different values of the number of paths per tunnel and the protection degree, i.e. $q$, for NKCP  and SCIP, respectively.
In the two figures, we observe that the number of solved instances is higher when the number of paths per tunnel is $6$. 
Indeed, with a high number of paths per tunnel, the traffic can be efficiently load-balanced in order to satisfy  link capacity constraints. Moreover, the higher the number of paths per tunnel is, the smaller  the bandwidth reservation is. Also of interest, the number of solved instances is much higher when $q=1$. This was expected since $q=2$ requires much more reserved bandwidth. Also, the mathematical models with $q=2$ are larger than with $q=1$. Therefore, a lot of instances are not solved due to time-limit. 

\begin{figure}[htbp]
    \centering 
    \includegraphics[width=1\columnwidth]{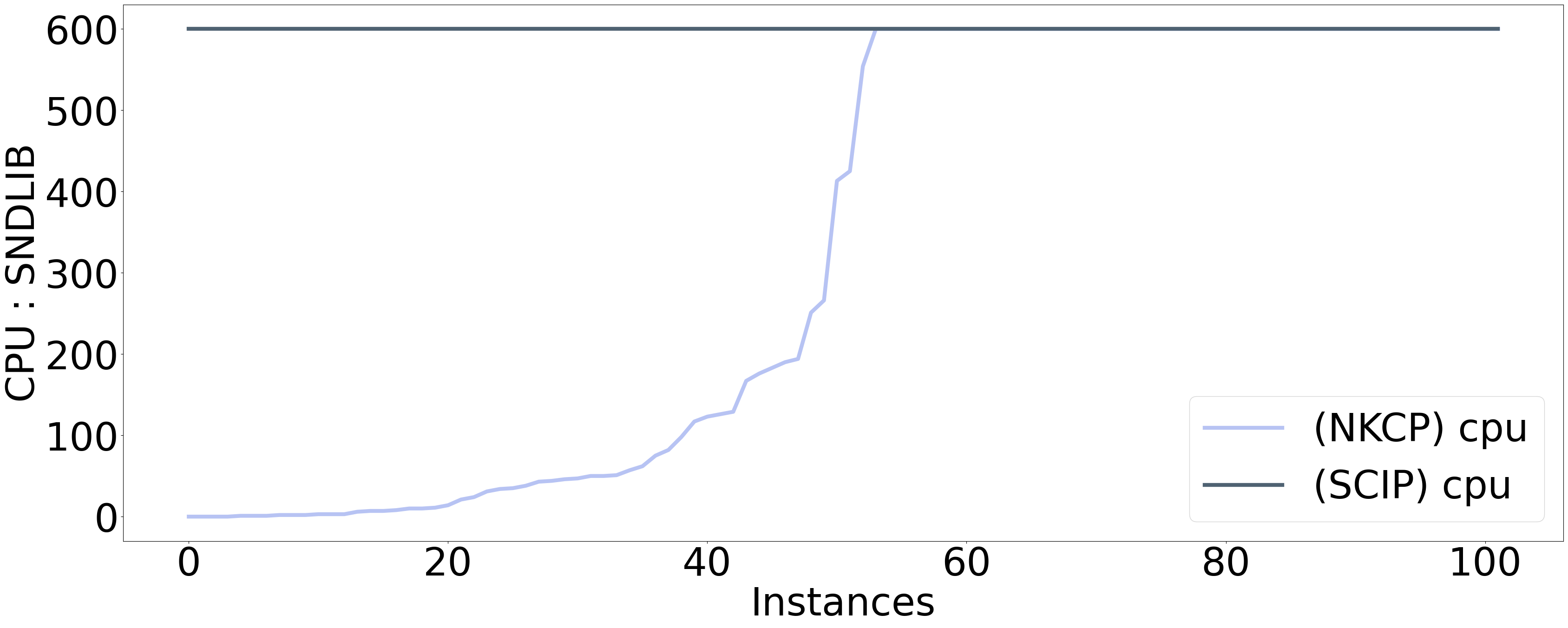}
    \caption{CPU times on SNDLIB instances.}
    \label{fig:cpu_sndlib}
\end{figure}

\begin{figure}[htbp]
    \centering 
    \includegraphics[width=1\columnwidth]{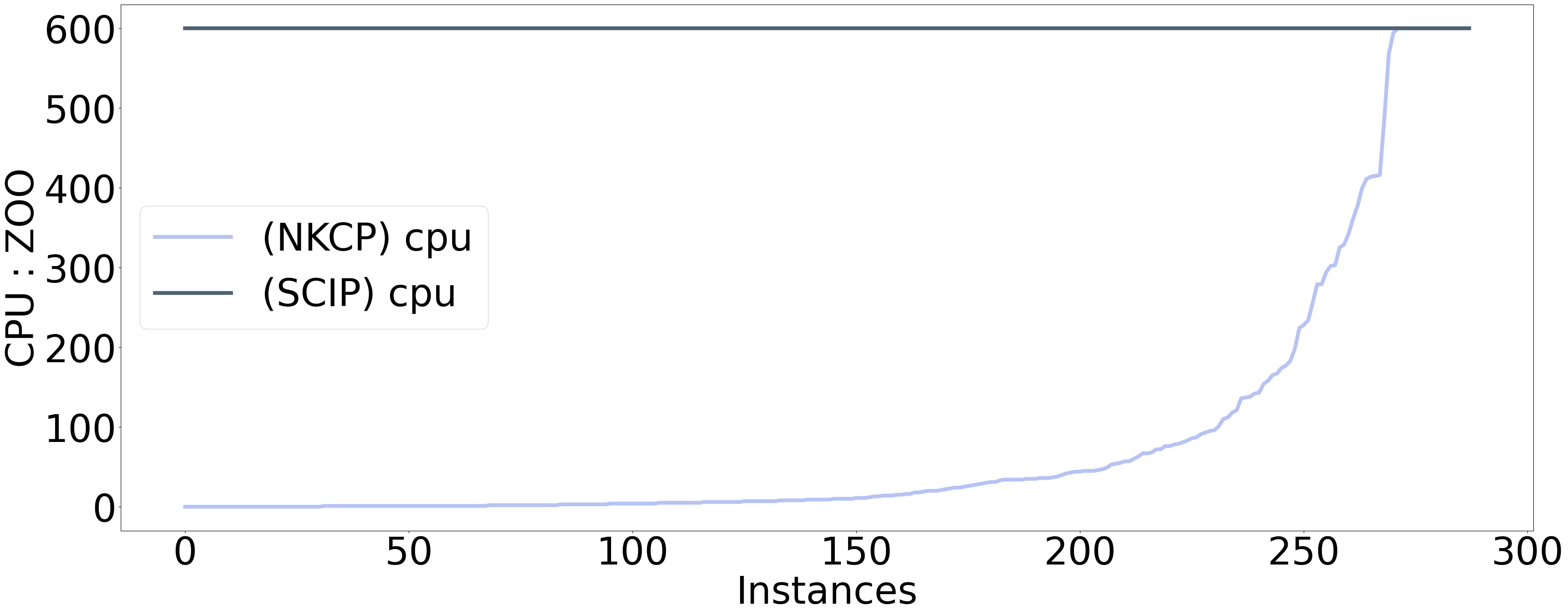} 
    \caption{CPU times on Internet Zoo topologies instances.}
    \label{fig:cpu_ZOO}
\end{figure}

Fig.~\ref{fig:cpu_sndlib} and \ref{fig:cpu_ZOO} compare the CPU times of SCIP and NKCP algorithms, respectively on SNDLIB and Internet Zoo instances. The instances are ranked in ascending order of CPU time associated to NKCP algorithm. We note that the NKCP algorithm solves almost all instances in a few seconds, while SCIP reaches the time-limit on most of the instances. Moreover, we observe that  SNDLIB instances are more difficult as NKCP reaches the time-limit multiple times. By comparing the two algorithms, we notice that NKCP solves more instances with 6 paths per tunnel and $q=2$.

\begin{figure}[t]
    \centering 
    \includegraphics[width=1\columnwidth]{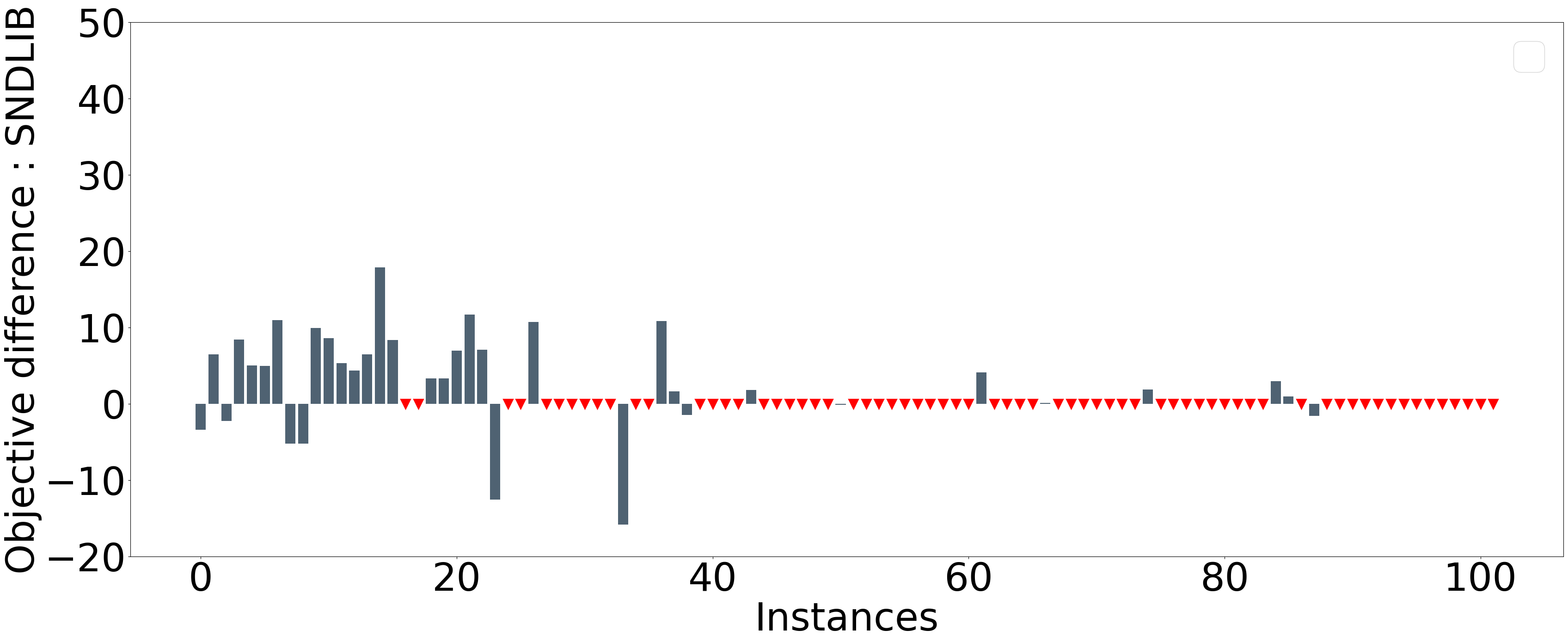} 
    \caption{Relative gap for bandwidth reservations between NKCP and SCIP on SNDLIB instances. Red markers indicate the unsolved instances by SCIP.}
    \label{fig:gap_sndlib_nkcp}
\end{figure}

\begin{figure}[t]
    \centering 
    \includegraphics[width=1\columnwidth]{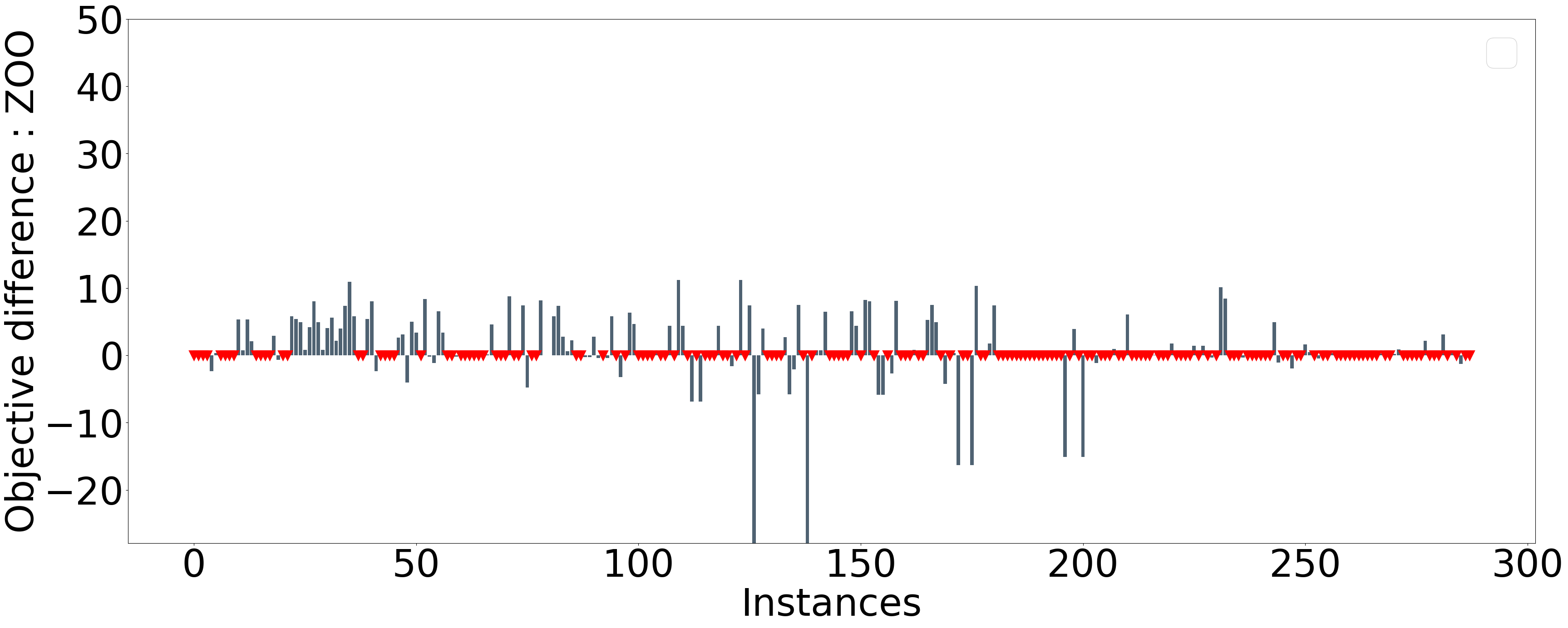}
    \caption{Relative gap of the bandwidth reservations given by NKCP and SCIP on Internet Zoo topologies instances. Red markers indicate the unsolved instances by SCIP.}
    \label{fig:gap_zoo_nkcp}
\end{figure}

Fig.~\ref{fig:gap_sndlib_nkcp} and \ref{fig:gap_zoo_nkcp} display the relative gap for bandwidth reservations between NKCP and SCIP, i.e., $\frac{(\text{obj}_{\text{NKCP}}-\text{obj}_{\text{SCIP}}) \times 100}{\text{obj}_{\text{SCIP}}}$, respectively for SNDLIB and Internet Zoo instances. Red markers represent the unsolved instances by SCIP. Note that all instances have been solved by NKCP. For both types of instances, we clearly see that SCIP fails to solve a lot of instances. 
On 90\% of the solved instances, we notice that the gap is lower than $10\%$. Due to the time-limit, we also notice that the gap is sometimes negative. Indeed, when SCIP finds a feasible solution  that it not optimal, the gap can be negative.   

NKCP is an iterative algorithm, at each iteration several cuts are added. In our experiments, NKCP generates around 27000 cuts on Internet Zoo topologies and 150000 on SNDLIB Instances. The difference is related to the number of links in the two types of instances. Indeed, in SNDLIB the number of links reaches 102 while it is 45 for Internet Zoo topologies. Hence, the number of SRLGs is much higher on SNDLIB leading to a higher number of Constraints \eqref{4} and \eqref{5}.

In the following, we show that using a neural network allows to approximate the non-convex function with a tight convex one. This leads to an efficient NKCP algorithm. To do so, we now compare two versions of NKCP, using different approximation methods: 1) a neural network and 2) a linear regression.  
In our case, we exploit the linear regression to obtain a linear function that approximates the non-linear non-convex function $\frac{x}{1-y}$. This latter is  $P(x,y)=1.299 x + 0.748 y -0.169$.

\begin{figure}[t]
\centering    
\includegraphics[scale=0.1]{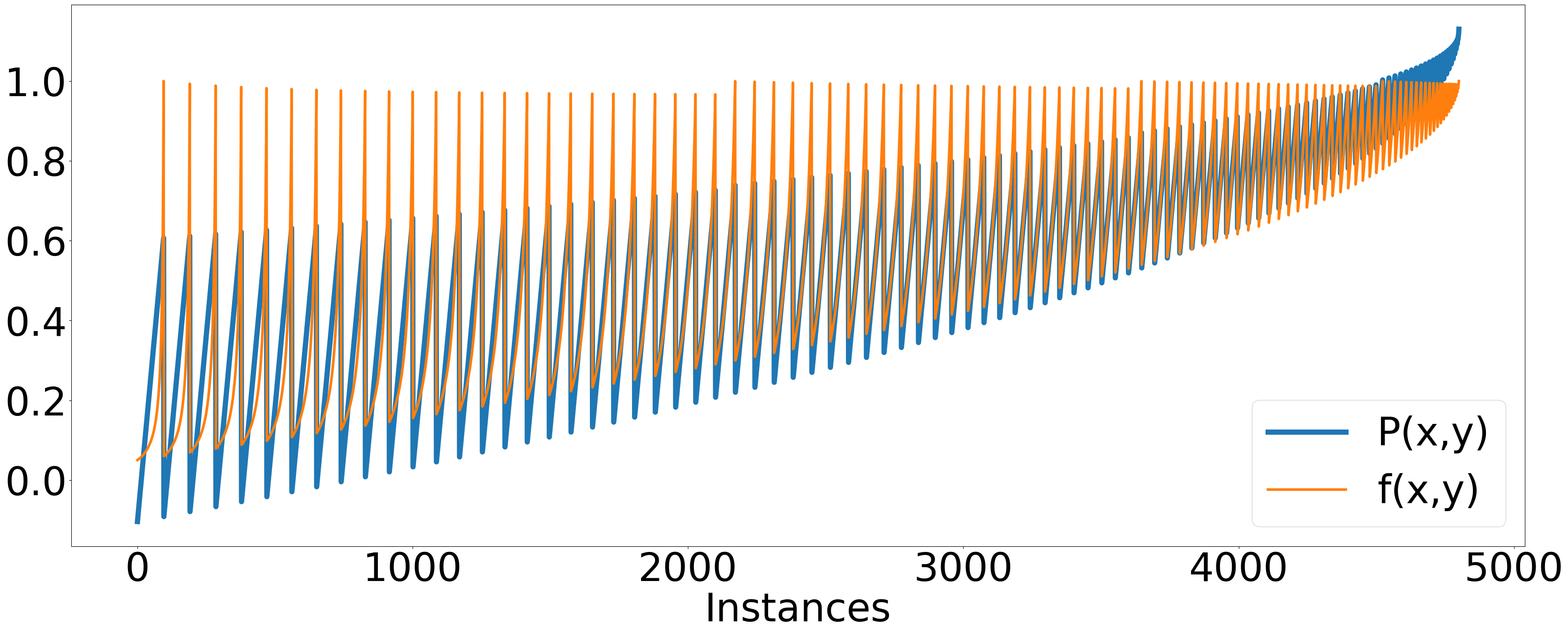} 
\caption{Comparison of the original function $f(x,y)$ and the approximation one $P(x,y)$ obtained with a linear
regression.}
\label{linear_regression}
\end{figure}

\begin{figure}[!b]
    \centering 
    \includegraphics[width=1\columnwidth]{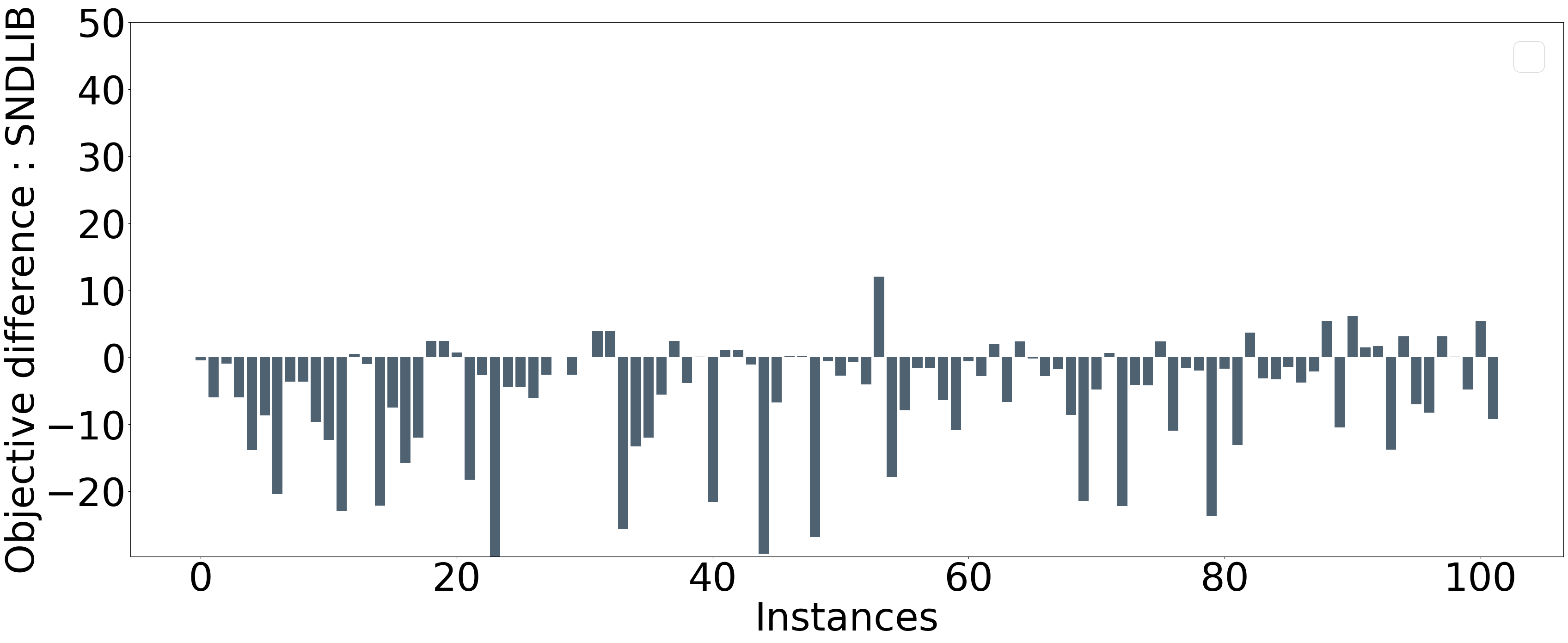}
    \caption{Relative gap of the bandwidth reservations given by NKCP and NKCP-R on SNDLIB instances.}
    \label{fig:obj_sndlib_nkcp}
\end{figure}

\begin{figure}[!b]
    \centering 
    \includegraphics[width=1\columnwidth]{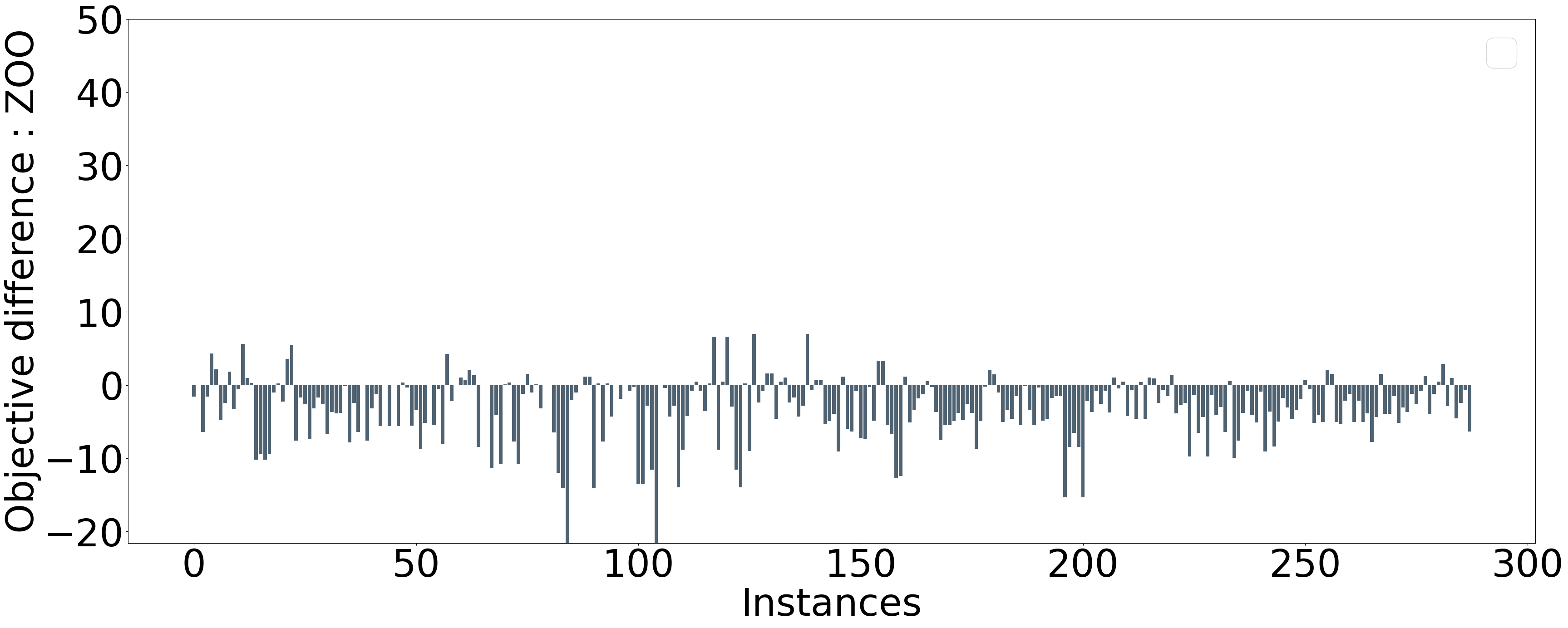}
    \caption{Relative gap of the bandwidth reservations given by NKCP and NKCP-R on Internet Zoo topologies instances.}
    \label{fig:obj_zoo_nkcp}
\end{figure}

Fig.~\ref{linear_regression} compares the original function $f(x,y)$ and the approximated one $P(x,y)$ on a set of points (the same as for the neural network in Sec.~\ref{sec:approximation_non_convex}). We clearly see that the approximation function under-approximates $f(x,y)=\frac{x}{1-y}$ on most instances (particularly those where $x$ is close to $0$) with a difference reaching $0.4$. On the other hand, it over-approximates it on instances where $x$ is close to $1.0$. Let denote by \textit{NKCP-R} the version of NKCP with linear regression method. Fig.~\ref{fig:obj_sndlib_nkcp} and~\ref{fig:obj_zoo_nkcp}
plot the relative gap for bandwidth reservations between NKCP-R and NKCP, i.e., $\frac{(\text{obj}_{\text{NKCP}}-\text{obj}_{\text{NKCP-R}}) }{\text{obj}_{\text{NKCP-R}}}$ for SNDLIB and Internet Zoo topologies instances, respectively. %, on the unshared and the shared variants. 
We can observe that the gaps reach $-15\%$ and even $-20\%$. On most instances, the gaps are negative, showing that a more accurate approximation is required. 
All instances are solved with NKCP and with NKCP-R.

\section{Conclusion}\label{sec:conclusion}
We have introduced a new protection mechanism for load balancing that makes an efficient use of bandwidth. To calculate "safe" split ratios, we  proposed a non-linear non-convex model and solved it using SCIP solver.
For the sake of performance, we proposed a novel method (NKCP) to solve a non-convex non-linear program using a linear solver. It consists in transforming the program into a convex non-linear one using a neural network and solve it using Kelley's cutting plane algorithm. Our numerical experiments showed very good performance for NKCP in terms of CPU time and objective value. Indeed, our results indicate that the NKCP improves, significantly, the CPU time compared to the compact model, without a significant loss on reservation cost. This new protected load balancing mechanism can operate on top of any load balancing solution as long as split ratios and routing paths can be controlled. It helps saving bandwidth compared to legacy methods that protect each  path individually. In the future, we investigate other non-convex function related to routing problem to generalize the approach.

\bibliographystyle{IEEEtran}
\bibliography{cas-refs}

\begin{thebibliography}{10}
\providecommand{\url}[1]{#1}
\csname url@samestyle\endcsname
\providecommand{\newblock}{\relax}
\providecommand{\bibinfo}[2]{#2}
\providecommand{\BIBentrySTDinterwordspacing}{\spaceskip=0pt\relax}
\providecommand{\BIBentryALTinterwordstretchfactor}{4}
\providecommand{\BIBentryALTinterwordspacing}{\spaceskip=\fontdimen2\font plus
\BIBentryALTinterwordstretchfactor\fontdimen3\font minus
  \fontdimen4\font\relax}
\providecommand{\BIBforeignlanguage}[2]{{%
\expandafter\ifx\csname l@#1\endcsname\relax
\typeout{** WARNING: IEEEtran.bst: No hyphenation pattern has been}%
\typeout{** loaded for the language `#1'. Using the pattern for}%
\typeout{** the default language instead.}%
\else
\language=\csname l@#1\endcsname
\fi
#2}}
\providecommand{\BIBdecl}{\relax}
\BIBdecl

\bibitem{paolucci2013survey}
F.~Paolucci, F.~Cugini, A.~Giorgetti, N.~Sambo, and P.~Castoldi, ``A survey on
  the path computation element (pce) architecture,'' \emph{IEEE Communications
  Surveys \& Tutorials}, vol.~15, no.~4, pp. 1819--1841, 2013.

\bibitem{medagliani2016global}
P.~Medagliani, J.~Leguay, M.~Abdullah, M.~Leconte, and S.~Paris, ``Global
  optimization for hash-based splitting,'' in \emph{2016 IEEE Global
  Communications Conference (GLOBECOM)}.\hskip 1em plus 0.5em minus 0.4em\relax
  IEEE, 2016, pp. 1--6.

\bibitem{zhou2014wcmp}
J.~Zhou, M.~Tewari, M.~Zhu, A.~Kabbani, L.~Poutievski, A.~Singh, and A.~Vahdat,
  ``Wcmp: Weighted cost multipathing for improved fairness in data centers,''
  in \emph{Proceedings of the Ninth European Conference on Computer Systems},
  2014, pp. 1--14.

\bibitem{rfc6790}
\BIBentryALTinterwordspacing
K.~Kompella, J.~Drake, S.~Amante, W.~Henderickx, and L.~Yong, ``{The Use of
  Entropy Labels in MPLS Forwarding},'' RFC 6790, Nov. 2012. [Online].
  Available: \url{https://www.rfc-editor.org/info/rfc6790}
\BIBentrySTDinterwordspacing

\bibitem{NARAGHIPOUR20082360}
M.~Naraghi-Pour and V.~Desai, ``Loop-free traffic engineering with path
  protection in mpls vpns,'' \emph{Computer Networks}, vol.~52, no.~12, pp.
  2360--2372, 2008.

\bibitem{zhou2000survivability}
D.~Zhou and S.~Subramaniam, ``Survivability in optical networks,'' \emph{IEEE
  network}, vol.~14, no.~6, pp. 16--23, 2000.

\bibitem{kamal2010overlay}
A.~E. Kamal, A.~Ramamoorthy, L.~Long, and S.~Li, ``Overlay protection against
  link failures using network coding,'' \emph{IEEE/ACM transactions on
  networking}, vol.~19, no.~4, pp. 1071--1084, 2010.

\bibitem{kelley1960cutting}
J.~E. Kelley, Jr, ``The cutting-plane method for solving convex programs,''
  \emph{Journal of the society for Industrial and Applied Mathematics}, vol.~8,
  no.~4, pp. 703--712, 1960.

\bibitem{GamrathEtal2020OO}
\BIBentryALTinterwordspacing
SCIP, ``The scip optimization suite 7.0,'' March 2020. [Online]. Available:
  \url{http://www.optimization-online.org/DB_HTML/2020/03/7705.html}
\BIBentrySTDinterwordspacing

\bibitem{mello2005dynamic}
D.~A. Mello, J.~U. Pelegrini, R.~P. Ribeiro, D.~A. Schupke, and H.~Waldman,
  ``Dynamic provisioning of shared-backup path protected connections with
  guaranteed availability requirements,'' in \emph{2nd International Conference
  on Broadband Networks, 2005.}\hskip 1em plus 0.5em minus 0.4em\relax IEEE,
  2005, pp. 1320--1327.

\bibitem{doucette2005advances}
J.~E. Doucette, ``Advances on design and analysis of mesh-restorable
  networks,'' Ph.D. dissertation, University of Alberta, 2005.

\bibitem{kang2015efficient}
N.~Kang, M.~Ghobadi, J.~Reumann, A.~Shraer, and J.~Rexford, ``Efficient traffic
  splitting on commodity switches,'' in \emph{Proceedings of the 11th ACM
  Conference on Emerging Networking Experiments and Technologies}, 2015, pp.
  1--13.

\bibitem{kang2014niagara}
N.~Kang, M.~Ghobadi, J.~Reumann, A.~Shraer, and R.~Jennifer, ``Niagara:
  Scalable load balancing on commodity switches,'' \emph{Princeton, NJ, USA,
  Tech. Rep. TR-973-14}, 2014.

\bibitem{yang2014keep}
B.~Yang, J.~Liu, S.~Shenker, J.~Li, and K.~Zheng, ``Keep forwarding: Towards
  k-link failure resilient routing,'' in \emph{IEEE INFOCOM 2014-IEEE
  Conference on Computer Communications}.\hskip 1em plus 0.5em minus
  0.4em\relax IEEE, 2014, pp. 1617--1625.

\bibitem{griva2009linear}
I.~Griva, S.~G. Nash, and A.~Sofer, \emph{Linear and nonlinear
  optimization}.\hskip 1em plus 0.5em minus 0.4em\relax Siam, 2009, vol. 108.

\bibitem{liberti2008introduction}
L.~Liberti, ``Introduction to global optimization,'' \emph{Ecole
  Polytechnique}, 2008.

\bibitem{lin2013review}
M.-H. Lin, J.~G. Carlsson, D.~Ge, J.~Shi, and J.-F. Tsai, ``A review of
  piecewise linearization methods,'' \emph{Mathematical problems in
  Engineering}, vol. 2013, 2013.

\bibitem{devore1998nonlinear}
R.~A. DeVore, ``Nonlinear approximation,'' \emph{Acta numerica}, vol.~7, pp.
  51--150, 1998.

\bibitem{csaji2001approximation}
B.~C. Cs{\'a}ji \emph{et~al.}, ``Approximation with artificial neural
  networks,'' \emph{Faculty of Sciences, Etvs Lornd University, Hungary},
  vol.~24, no.~48, p.~7, 2001.

\bibitem{bertsekas1997nonlinear}
D.~P. Bertsekas, ``Nonlinear programming,'' \emph{Journal of the OR Society},
  vol.~48, no.~3, pp. 334--334, 1997.

\bibitem{kiwiel2007lagrangian}
K.~C. Kiwiel, T.~Larsson, and P.~O. Lindberg, ``Lagrangian relaxation via
  ballstep subgradient methods,'' \emph{Mathematics of Operations Research},
  vol.~32, no.~3, pp. 669--686, 2007.

\bibitem{nesterov1994interior}
Y.~Nesterov and A.~Nemirovskii, \emph{Interior-point polynomial algorithms in
  convex programming}.\hskip 1em plus 0.5em minus 0.4em\relax SIAM, 1994.

\bibitem{duran1986outer}
M.~A. Duran and I.~E. Grossmann, ``An outer-approximation algorithm for a class
  of mixed-integer nonlinear programs,'' \emph{Mathematical programming},
  vol.~36, no.~3, pp. 307--339, 1986.

\bibitem{grotschel1981ellipsoid}
M.~Gr{\"o}tschel, L.~Lov{\'a}sz, and A.~Schrijver, ``The ellipsoid method and
  its consequences in combinatorial optimization,'' \emph{Combinatorica},
  vol.~1, no.~2, pp. 169--197, 1981.

\bibitem{westerlund1995extended}
T.~Westerlund and F.~Pettersson, ``An extended cutting plane method for solving
  convex minlp problems,'' \emph{Computers \& chemical engineering}, vol.~19,
  pp. 131--136, 1995.

\bibitem{desaulniers2006column}
G.~Desaulniers, J.~Desrosiers, and M.~M. Solomon, \emph{Column
  generation}.\hskip 1em plus 0.5em minus 0.4em\relax Springer Science \&
  Business Media, 2006, vol.~5.

\bibitem{cplex}
\BIBentryALTinterwordspacing
{IBM}, ``{ILOG CPLEX Solver}.'' [Online]. Available:
  \url{https://www.ibm.com/analytics/cplex-optimizer}
\BIBentrySTDinterwordspacing

\bibitem{chollet2015keras}
F.~Chollet \emph{et~al.}, ``Keras,'' \url{https://keras.io}, 2015.

\bibitem{scikit-learn}
F.~Pedregosa, G.~Varoquaux, A.~Gramfort, V.~Michel, B.~Thirion, O.~Grisel,
  M.~Blondel, P.~Prettenhofer, R.~Weiss, V.~Dubourg, J.~Vanderplas, A.~Passos,
  D.~Cournapeau, M.~Brucher, M.~Perrot, and E.~Duchesnay, ``Scikit-learn:
  Machine learning in {P}ython,'' \emph{Journal of Machine Learning Research},
  vol.~12, pp. 2825--2830, 2011.

\bibitem{zooTopology}
S.~Knight, H.~Nguyen, N.~Falkner, R.~Bowden, and M.~Roughan, ``The internet
  topology zoo,'' \emph{Selected Areas in Communications, IEEE Journal on},
  vol.~29, no.~9, pp. 1765 --1775, october 2011.

\bibitem{SNDlib10}
S.~Orlowski, M.~Pi{\'o}ro, A.~Tomaszewski, and R.~Wess{\"a}ly,
  ``\BIBforeignlanguage{English}{{SNDlib} 1.0--{S}urvivable {N}etwork {D}esign
  {L}ibrary},'' in \emph{\BIBforeignlanguage{English}{3rd International Network
  Optimization Conference}}, April 2007.

\end{thebibliography}

\end{document}